\documentclass[fleqn,10pt]{wlscirep}
\usepackage[utf8]{inputenc}
\usepackage[T1]{fontenc}
\usepackage{lipsum} 
\title{Imitation vs Serendipity in Ranking Dynamics}

\author[1,2]{Federica De Domenico}
\author[3,4,5]{Fabio Caccioli}
\author[1,3]{Giacomo Livan}
\author[1,2]{Guido Montagna}
\author[2]{Oreste Nicrosini}
\affil[1]{Dipartimento di Fisica, Universit\`a degli Studi di Pavia, Via A. Bassi 6, 27100, Pavia, Italy}
\affil[2]{Istituto Nazionale di Fisica Nucleare, Sezione di Pavia, Via A. Bassi 6, 27100, Pavia, Italy}
\affil[3]{Department of Computer Science, University College London, 66-72 Gower Street, London WC1E 6EA, United Kingdom}
\affil[4]{Systemic Risk Centre, London School of Economics and Political Science, London WC2A 2AE, United Kingdom}
\affil[5]{London Mathematical Laboratory, 8 Margravine Gardens, London WC 8RH, United Kingdom}

\begin{abstract}
Participants in socio-economic systems are often ranked based on their performance. Rankings conveniently reduce the complexity of such systems to ordered lists. Yet, it has been shown in many contexts that those who reach the top are not necessarily the most talented, as chance plays a role in shaping rankings. Nevertheless, the role played by chance in determining success, i.e., serendipity, is underestimated, and top performers are often imitated by others under the assumption that adopting their strategies will lead to equivalent results. We investigate the tradeoff between imitation and serendipity in an agent-based model. Agents in the model receive payoffs based on their actions and may switch to different actions by either imitating others or through random selection. When imitation prevails, most agents coordinate on a single action, leading to non-meritocratic outcomes, as a minority of them accumulates the majority of payoffs. Yet, such agents are not necessarily the most skilled ones. When serendipity dominates, instead, we observe more egalitarian outcomes. The two regimes are separated by a sharp transition, which we characterise analytically in a simplified setting. We discuss the implications of our findings in a variety of contexts, ranging from academic research to business.
\end{abstract}
\begin{document}

\flushbottom
\maketitle
\thispagestyle{empty}
\section*{Introduction}
Rankings of individuals, companies and institutions based on their performance have become ubiquitous, reducing complex systems to ordered lists that reflect the ability of their participants to perform precise functions~\cite{rankingmobility2}. Rankings are used in a variety of domains, ranging from natural to social, from economic to infrastructural ones~\cite{erdi2019ranking}, with the aim of a better allocation of resources, funds and rewards~\cite{rankingmobility1}. The emphasis on rankings has spurred considerable interest in their dynamics over time \cite{rankingmobility3}, leading to the birth of a novel research field named ``ranking of rankings'', which investigates the goodness of a ranking's evaluation criteria~\cite{rankingofranking}. 

One of the main functions of rankings is to help identifying the strategies of top performers, which are often recognised as the best practices that others should adopt. In fact, individuals, organisations and institutions often assume that they will be equally successful if they imitate top performers' strategies. However, there is considerable evidence that such an approach can sometimes backfire and result in a weak association between skills and measured performance~\cite{performanceability}, and that agents can be more successful when they develop their own strategies~\cite{Giacomo2019} or pursue risky ones~\cite{guedj2005experts,performanceability} rather than imitate those of others. 

Imitation in ranking dynamics is often grounded in social influence, which often drives individuals' decision-making and shapes the collective wisdom of the crowd~\cite{socialinfluence, wisdomofthecrowd,bond1996culture,cialdini2004social}. A prime  example of these effects was illustrated in a much celebrated experiment in an artificial music market, which resulted in very low correlations between a song's success when social interactions between market participants were swtiched on/off~\cite{artificialmarket}. 

The role played by chance in successful paths is often underestimated. For instance, in~\cite{talentvsluck,challet2020origins} it was shown that in a synthetic society of agents luck prevails over talent in favouring an agent's rise to the top. At the same time, chance rarely compensates over time for self-reinforcing mechanisms usually referred to as the ``rich-get-richer'' effect or the Matthew effect~\cite{mattheweffect,petersen2014reputation}, according to which success breeds more success, often to the point that a (sometimes random) early competitive advantage can lead to long-lasting consequences~\cite{li2019early}. The interplay between chance and those effects has been extensively investigated in the literature. Recently, and agent-based models has shown that, provided that some degree of skill is necessary to be successful in life, it is rare that the most talented individuals reach the highest peaks of success, while averagely talented but sensibly luckier individuals reach the top of rankings~\cite{talentvsluck}.  

In this paper we explore the consequences of imitation versus chance in an artificial society whose agents are ranked based on a notion of performance. After characterising the agents' intrinsic skills, we analyse the payoff gained by the society under different scenarios, and show which of those lead to more/less meritocratic outcomes. Our results capture quite nicely some of the aforementioned phenomena. When imitation prevails, the top performers are not the most skilled ones and, at the same time, the most skilled agents are not the ones who accumulate the highest payoffs. On the contrary, when chance is the dominant mechanism, society becomes more meritocratic. In the latter scenario, we can speak of serendipity, namely positive developments of events that occur in an unplanned manner~\cite{YAQUB2018169}.

\section*{Results}
\subsection*{Model implementation}
\begin{figure}
\centering
\includegraphics[width=0.6\linewidth]{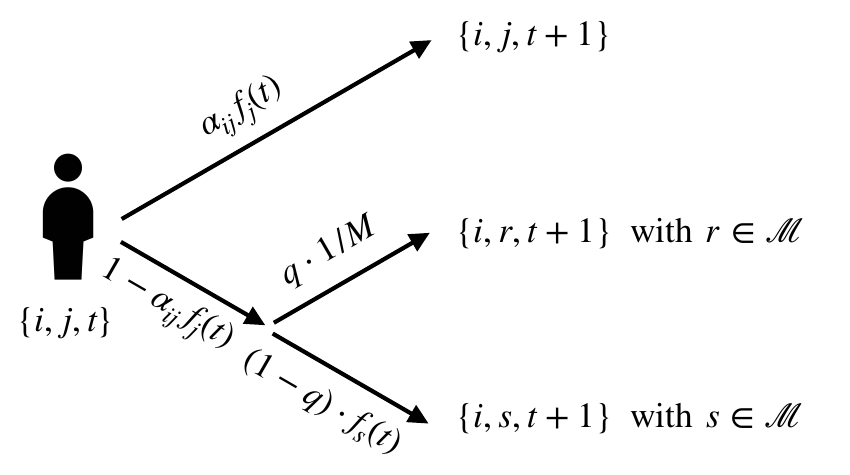}
\caption{Representation of the model dynamics for a generic agent $i$ who is in state $j$ at time $t$. All the possible transition outcomes are reported with their probability. Parameter $\alpha_{ij}$ represents the expertise of the agent $i$ in field $j$, $f_j(t)$ is the fraction of agents employed in the activity $j$ at time $t$, $M$ is the number of possible actions and the parameter $q$ accounts for randomness in the switching.}
\label{fig:dinamica}
\end{figure}

We consider an agent-based model with a discrete-time dynamics, in which agents have to decide at each time step whether to persist with their current action or switch to a different one. It is conceived as a memory-less process since each move is independent of the previous ones.  A schematic representation of the dynamics is shown in Fig.~\ref{fig:dinamica}. More in detail, there are $N$ agents, labelled with indices $i \in \mathcal{N} = \{1,...,\, N\}$, who are empowered to take $M$ actions, labelled with indices $j \in \mathcal{M}= \{1,...,\, M\}$. In the following, we will write $j(i,t)$ to denote the fact that agent $i$ plays action $j$ at time $t$. In order to keep our notation as light as possible, we will simply indicate $j(i,t)$ as $j$ whenever possible.  Each action $j$ is associated with a societal impact $\pi_j \in [0,1]$. Actions with high societal impact ($\pi_j\rightarrow 1$) have beneficial effects on society as a whole, whereas actions with low societal impact ($\pi_j\rightarrow 0$) do not. For instance, in the context of academic research, we may think of an action characterised by a high $\pi_j$ as a research field such as, e.g., cancer research, whereas an action with low $\pi_j$ would correspond to a niche field.

As mentioned above, agents in the model may switch to a different action at any given time step. The switching mechanism is not trivial and depends on a few factors. Among these we find an agent's skills, defined through a random matrix whose elements are denoted with $\alpha_{ij} \in [0,1]$. A high value of $\alpha_{ij}$ indicates that agent $i$ is skilled at action $j$, a low value indicates the opposite. Note that both $\alpha_{ij}$ and $\pi_j$ are randomly extracted from a uniform distribution in $[0,1]$ and are assigned at the beginning of each simulation. Another relevant quantity is the appeal of each action, represented by the fraction of agents $f_{j}(t)$ who are performing action $j$ at time $t$. We identify an action's appeal with its popularity in order to encapsulate some of the mechanism discussed in the Introduction, namely the effects of imitation and social influence on individual choices.

The model's dynamics work as follows. At time $t=0$ agents are allocated to the available actions with probability $\pi_j$, i.e., proportionally to the actions' societal impact. At each time step $t = 0, ..., T$ each agent receives a payoff $\alpha_{ij} \, f_{j}(t)$. This assumption reflects the fact that agents will receive higher payoffs when they play actions that they are good at and/or that are popular, regardless of their societal impact. In fact, a large $f_j(t)$ may compensate for a relatively low $\alpha_{ij}$. Let us note that the payoff lies in $[0,1]$.

At $t' = t+1$ (with $t=0,..., T-1$) agents persevere with the action of time $t$ with probability equal to the previous time step's payoff, i.e., $\alpha_{ij} \, f_{j}(t)$, reflecting the incentive to stick with actions that are rewarding. In the Appendix, we implement a slightly different dynamics, for which agents persevere in their current action with probability $\alpha_{ij} \, f_j(t) \, \pi_j$, i.e., by taking into account an action's societal impact as well. 

In the baseline scenario, an agent will choose to switch actions with probability $1-  \alpha_{ij} \, f_{j}(t) $. In this case, an additional parameter $q \in [0,1]$ is introduced to interpolate between imitation and chance. Namely, with probability $q$ agents pick a new action at random. With complementary probability $1-q$ they pick their new action with probability $f_j(t)$, i.e., proportionally to its current popularity.

In this respect, $q$ quantifies to what extent agents pay attention to the rankings induced by the accumulation of payoffs (see next Section) and choose to imitate the actions of their peers~\cite{Giacomo2019}: when $q\rightarrow 0$, imitation becomes the dominant mechanism, while if $q\rightarrow 1$ agents randomly explore the space of available actions.

We implement simulations of our model for several values of $q$ in order to achieve a comprehensive understanding of the different scenarios. For each parameter set, we run $S=1000$ simulations to ensure robust statistical analysis, and we let each simulation run for $T= 500$ time steps, for which a steady or absorbing state (depending on the value) is reached.  We checked that outcomes of the simulations do not vary when we keep the ration $N/M$ fixed, so in the following we will use $N = 250$ agents and $M=100$ actions. The rationale behind the choice of $N>M$ is that there are likely more agents than possible actions to play in most realistic settings. Anyway, the qualitative outcomes of the model for $N \lessgtr M$ are the same unless explicitly mentioned. Error bars in the figures represent sample standard deviation.

\subsection*{Success and social disparity}
\begin{figure}
\centering
\includegraphics[width=0.45\linewidth]{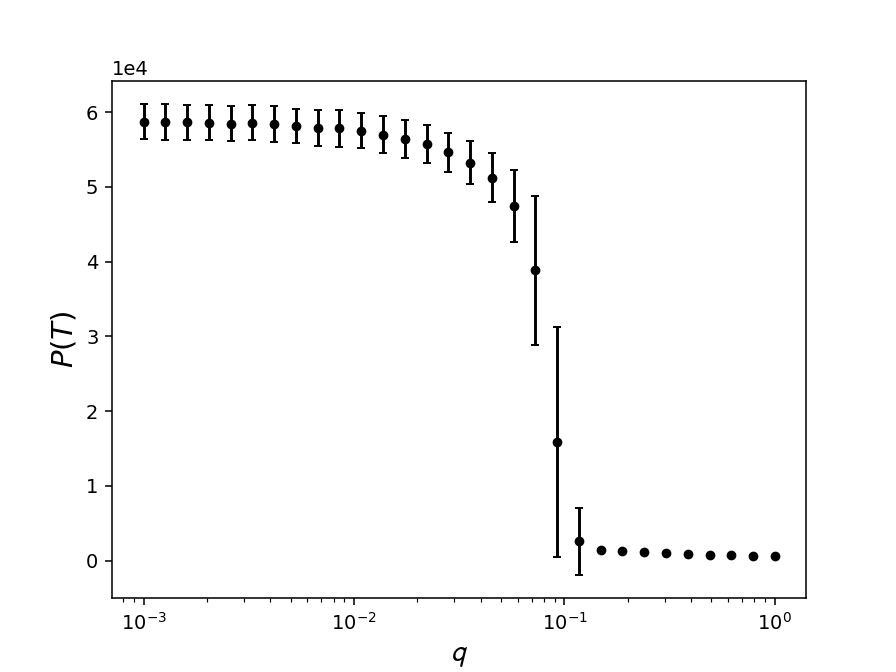} $\qquad \qquad$ \includegraphics[width=0.45\linewidth]{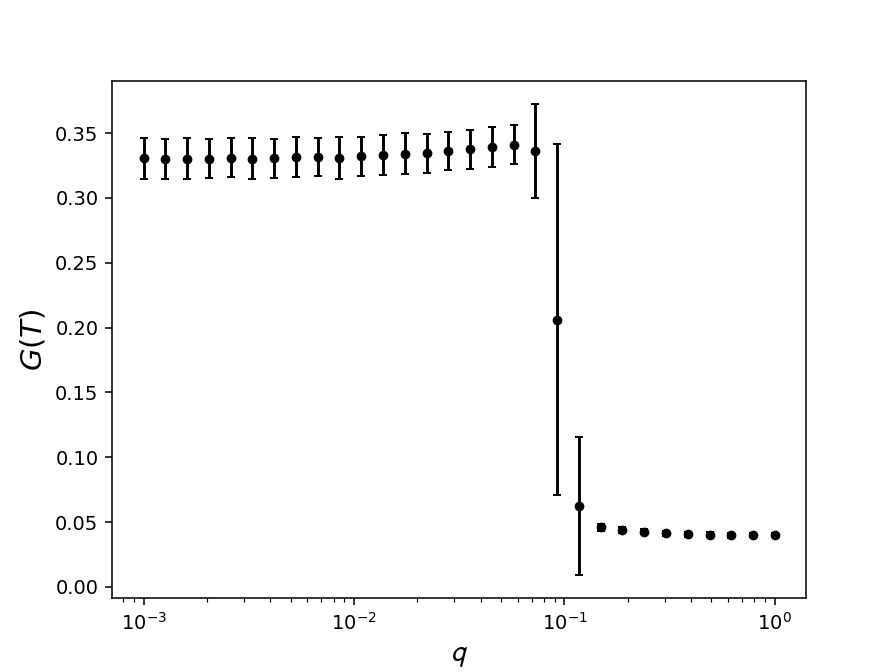}
\caption{Left panel: Cumulative societal payoff $P(T)$ for several values of parameter $q$. Right panel: Gini coefficient $G(T)$ for same values of $q$. Simulations have been implemented with $N=250$ agents, $M=100$ actions and $T=500$ time steps. Results are an average over a sample of size $S=1000$.}
\label{fig:payoff-gini}
\end{figure}
To investigate the outcome(s) of different scenarios, it is necessary to introduce some observables. First, we define the cumulative payoff that agent $i$ gains with their actions over time, from $t=0$ to $t=T$, as
\begin{equation}
    P_i(T)=\sum_{t=0}^{T} \alpha_{ij}\, f_j(t) \ .
\end{equation}
We remind the reader that in the above and the following equations $j = j(i,t)$, i.e., the action played by agent $i$ at time $t$. The overall societal payoff collectively generated by all agents is thus given by
\begin{equation}
    P(T)=\sum_{i=1}^{N} \, \sum_{t=0}^{T} \alpha_{ij}\, f_j(t) \ .
\end{equation}
We combine the analysis of $P(T)$ with the study of the Gini coefficient, a measure of inequality in a society, which is defined as
\begin{equation}
    G(T) = \frac{1}{N\, P(T)}\, \sum_{i<k} | P_i(T)-P_k(T)| \ , 
\end{equation}
leading to $G(T) = 0$ for perfect equality and $G(T) \rightarrow 1$ when cumulative payoffs are concentrated in the hands of very few agents. The results of these investigations are shown in Fig.~\ref{fig:payoff-gini}. 
When imitation is the prevailing mechanism, i.e. $q\rightarrow 0$, the overall payoff $P(T)$ reaches its maximum, though this happens at a cost. In fact, the Gini coefficient $G(T)$ shows that this corresponds to an unequal society, in which the majority of the payoff is generated by a minority of the agents. On the contrary, when randomness dominates (for $q \rightarrow 1$), we observe a lower societal payoff distributed more equally. Between these two limiting conditions, a sharp transition happens for a certain value $q^*$, for which we provide some intuition in a simplified setting in a following section. Note that error bars are particularly large at the phase transition, in line with behaviours observed for phase transitions in condensed matter physics.

\begin{figure}
\centering
\includegraphics[width=0.45\linewidth]{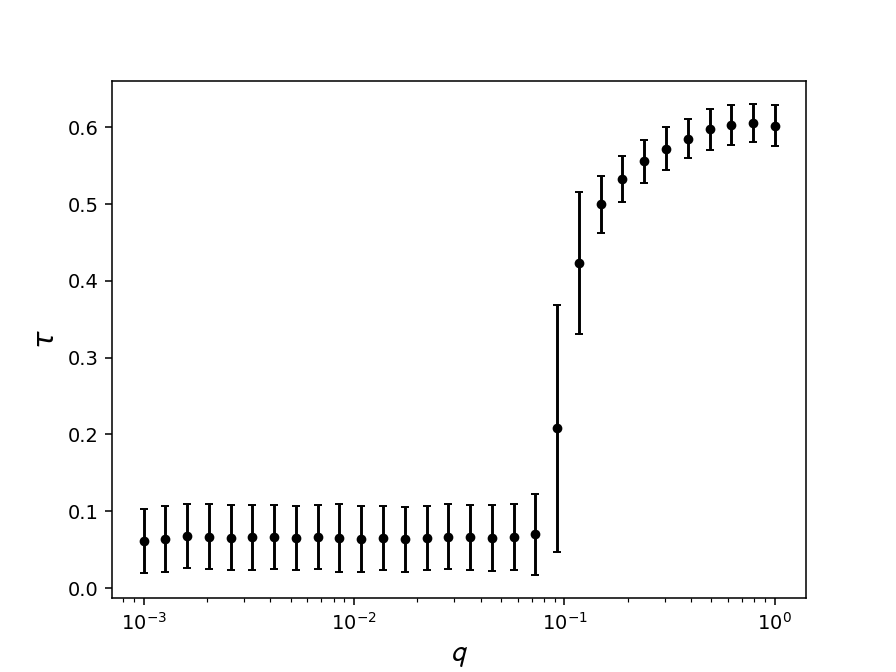} $\qquad \qquad$
\includegraphics[width=0.45\linewidth]{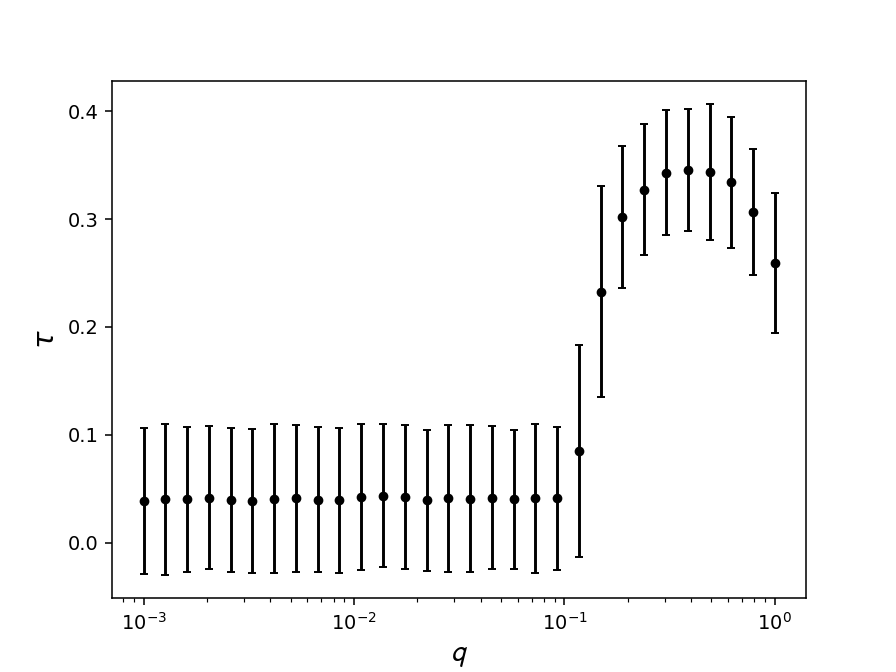}
\caption{Mean Kendall rank correlation coefficient $\tau$ between $P_i(T)$ and $\phi_i^{avg}$. Left panel: $N=250,\,M=100$. Right panel: $N=100, \,M=250$.  Other simulation parameters are the same of Fig.~\ref{fig:payoff-gini}.}
\label{fig:correlation}
\end{figure}

It is fair to wonder which of the two above societies represents a better outcome: a ``richer'' but unequal one or a ``poorer'' but more egalitarian one? Intuitively, the former would be preferable when higher individual payoffs reflect an agent's superior skills, i.e., when outcomes are meritocratic. To this aim, we compute the Kendall rank correlation coefficient $\tau$ --- a measure of similarity between two rankings~\cite{kendall} --- between the agents' cumulative payoff $P_i(T)$ and intrinsic fitness. We define the latter as the average skill of an agent across all possible actions:
\begin{equation}
    \phi_i^{avg} = \frac{1}{M}\, \sum_{j=1}^M \alpha_{ij}.
\end{equation}
As shown in Fig.~\ref{fig:correlation}, for $q<q^*$ there is no correlation between the agents' payoffs and fitness. As $q$ increases, we find again a sharp transition at $q=q^*$ and, for $q>q^*$, some correlation between payoffs and fitness (i.e., ultimately, meritocracy) is restored. This behaviour changes slightly depending on the ratio $N/M$. When $M>N$, $\tau$ behaves non monotonically; when $N>M$, instead, it increases monotonically. In line with~\cite{talentvsluck}, the above results further confirm that the most successful agents (i.e., those who accumulate the highest payoff) in general are not the most skilled ones.

Combining the different elements, these analyses show that imitation could be the winning strategy only for a limited number of agents and that the most skilled ones would not be the most successful. On the contrary, when chance dominates, outcomes are serendipitous and society becomes more egalitarian.

\subsection*{Condensation and resource allocation}
This subsection is dedicated to the investigation of the distribution of the actions played by agents. In particular, we focus on the participation ratio, a measure of localization borrowed from condensed matter and quantum physics, defined as
\begin{equation}
PR(t)= \frac{1}{M} \frac{1}{\sum_{j} {\bigl (\tilde{f}_j}(t)\bigl)^4} 
\end{equation}
with $j \in \mathcal{M}$ and where $\tilde{f}_j(t)$ is the normalized version of the quantity $f_j (t)$ in $L^2$ norm and represents the fraction of agents who play action $j$ at time $t$. 

\begin{figure}
\centering
\includegraphics[width=0.45\linewidth]{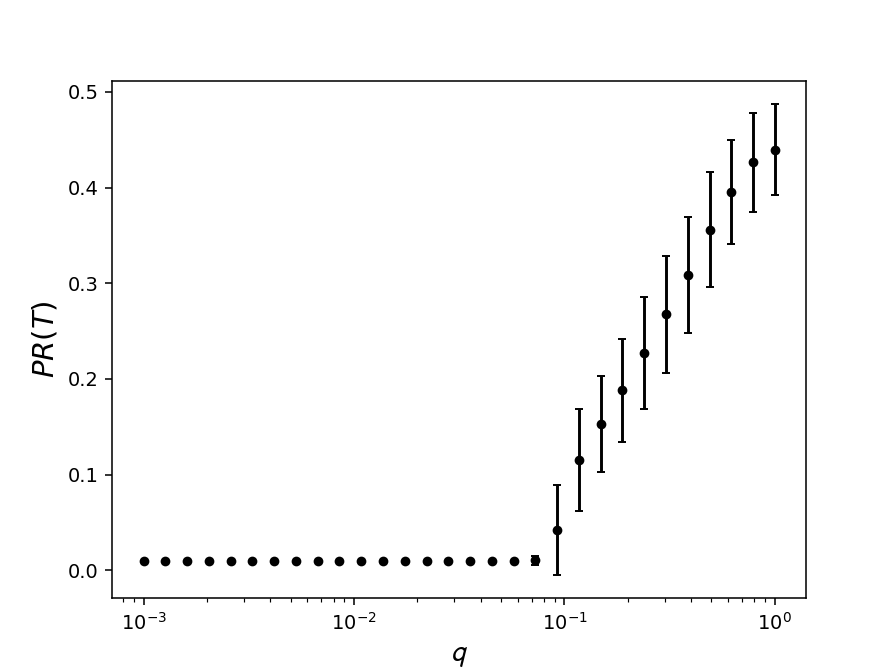} $\qquad \qquad$ \includegraphics[width=0.45\linewidth]{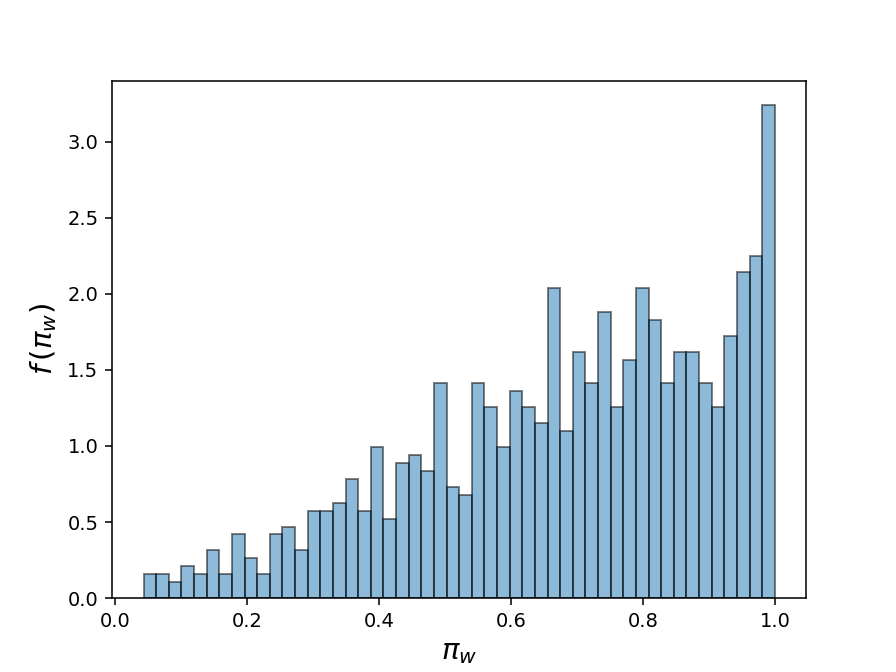} 
\caption{Left panel: Participation ratio $PR(T)$. Right panel: normalized distribution of the payoffs of the most played actions at time $T$ for $q = 0$. All parameters are the same of Fig.~\ref{fig:payoff-gini}.}
\label{fig:ipr}
\end{figure}

We focus our attention on $PR(T)$. As shown in Fig.~\ref{fig:ipr}, for $q<q^*$ all agents condensate on a single action, namely $\tilde{f}_s(T) =1$ for a certain $s \in \mathcal{M}$ and $f_i(T)=0, \forall i \neq s$. The activity $s$ resembles an absorbing state, as agents can temporarily leave it through the switching mechanism (when $q > 0$), but are bound to return to it through the very same mechanism. After the transition at $q=q^*$, other actions start to be populated. Yet, the uniform scenario $\tilde{f}_j(T) = 1/\sqrt{M}\,, \forall j \in \mathcal{M}$ (which is the other extremal condition) is never reached. We further elaborate on this point in the section devoted to analytical considerations below.

Condensation in a single action is a positive outcome when such action is beneficial to society. For that reason, we study the distribution across simulations of the societal payoffs for the most played action, which we refer to as the ``winning'' one, when $q=0$.  As we can see from the right panel of Fig.~\ref{fig:ipr}, in the majority of simulations the winning activity is associated with an above average societal payoff ($\pi_w > 1/2$). Yet, in a sizeable minority of cases ($\sim$ 20\% of simulations) it is not unlikely for the agents to condensate on an action with a below average societal payoff as a consequence of imitation. It is remarkable that the same scenario occurs even when the societal payoff $\pi_j$ is explicitly accounted for in the switching mechanism, namely agents keep playing the same action with probability proportional to $\alpha_{ij}\, f_j(t)\, \pi_j$, as shown in Fig.~\ref{fig:benefit_app}.

Lastly, we aim to address the following question: are the agents playing actions that are beneficial to society good at them? To this end, we introduce a simple measure of societal benefit
\begin{equation}
    B(t) = \sum_{i=1}^N\,\alpha_{ij} \, \pi_j \ .
\end{equation}
whose cumulative value over time is given by $B(T) = \sum_{t=0}^T B(t)$. Let us remind here that $j$ here is short for $j(i,t)$, i.e., the action undertaken by agent $i$ at time step $t$. Higher values of $B(T)$ indicate that the model's dynamics naturally incentivises agents to play actions that generate societal benefits when they are good at them. 

\begin{figure}
\centering
\includegraphics[width=0.45\linewidth]{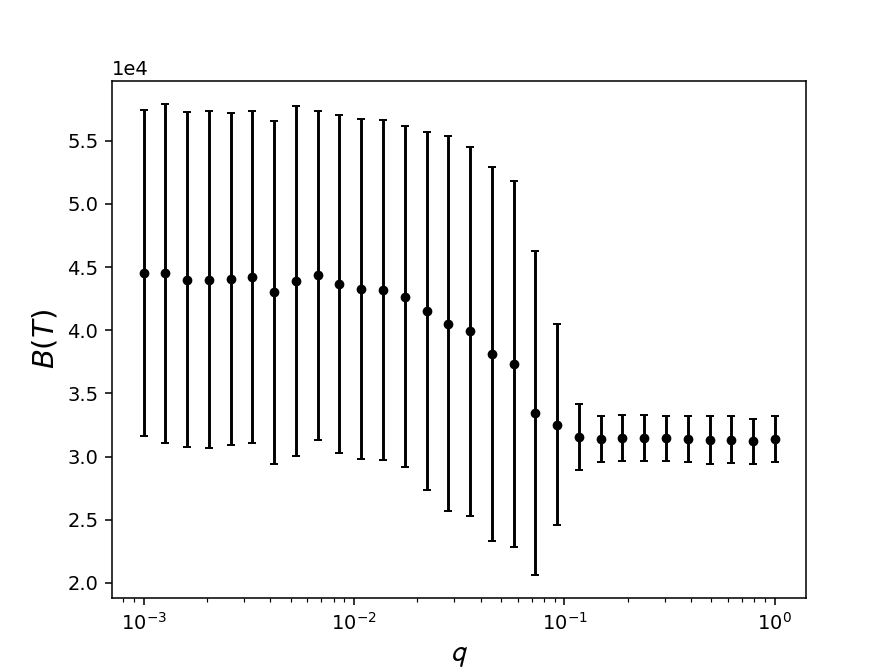}
\caption{The overall benefit $B(T)$ gained by the community due to agents' abilities in socially beneficial actions. Simulation parameters are the same of the previous analyses.}
\label{fig:benefit}
\end{figure}

As shown in Fig.~\ref{fig:benefit}, on average $B(T)$ is higher for $q < q^*$, yet its standard deviation is particularly large in this regime. After the transition, $q > q^*$ societal benefit decreases, together with its standard deviation.

\subsection*{Analytical considerations}
In order to gain some analytical insights on the model’s behaviour, we consider a simplified version with  $M = 2$ activities and $\alpha_{ij} = \alpha$ ($\forall i,\, j = 1, 2$). We define $n_j(t)$ as the number of agents who play action $j$ at time $t$, so that $n_1(t)+n_2(t) = N$. In addition, we replace the model’s probabilistic dynamics with deterministic transition rates. According to the dynamics shown in Fig.~\ref{fig:dinamica}, we study the flux of agents who enter and exit from activity $1$ at time $t'=t+1$. Focusing on agents who play action $1$ at time $t$, at $t'=t+1$ there are
\begin{itemize}
    \item[a)] $\alpha\, f_1(t)\, n_1(t)$ agents who at time $t$ are taking action $1$ and choose to keep playing it;
    \item[b)] $(1- \alpha\, f_1(t))\, n_1(t)$ agents who consider leaving activity $1$. Out of this group, $q\, (1- \alpha\, f_1(t))\, n_1(t)/2$ agents keep action $1$ as a consequence of random selection and $(1-q) \, f_1(t)\, (1- \alpha\, f_1(t))\, n_1(t)$ agents pick action $1$ because of its popularity.
\end{itemize}
Correspondingly, examining agents who are in action $2$ at time $t$, we have a total of $(1- \alpha\, f_2(t))\, n_2(t)$ agents who consider leaving it. Out of these, $q\, (1- \alpha\, f_2(t))\, n_2(t)/2$ agents end up in action $1$ through random selection and $(1-q) \, f_1(t)\, (1- \alpha\, f_2(t))\, n_2(t)$ agents end up in it because of its popularity.

\begin{figure}
\centering
\includegraphics[width=0.45\linewidth]{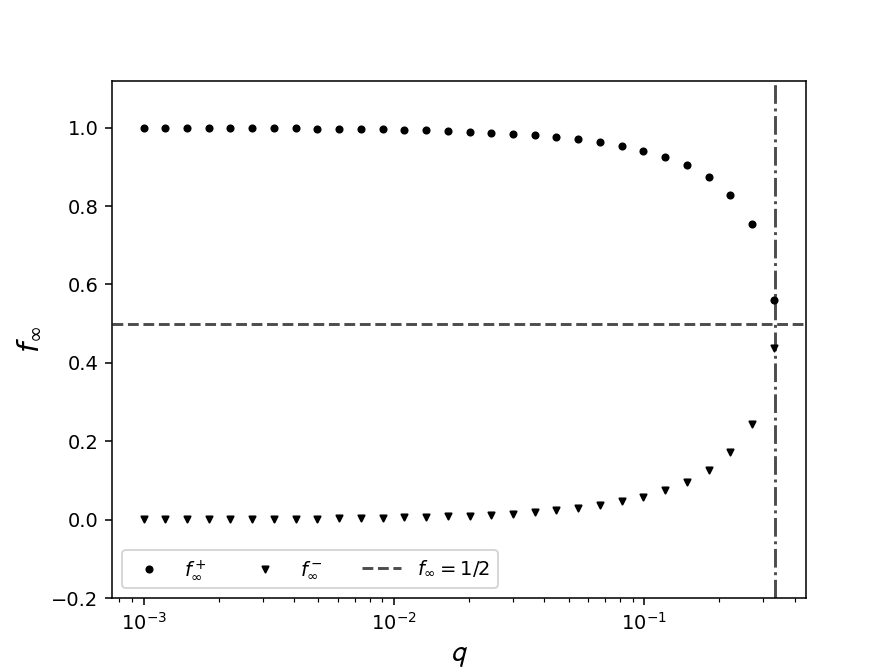}
\caption{Fraction of agents $f_\infty$ as a steady-state solution in a simplified scenario with two available actions. The horizontal line marks the trivial solution $f_\infty = 1/2$, while the vertical one underlines the value $q=q^*$ for which a sharp transition occurs.}
\label{fig:analytical}
\end{figure}

Considering the relations $n_2(t)= N-n_1(t)$ and $f_2(t)=1-f_1(t)$, we combine the previous contributions and write:
\begin{equation}
    n_1(t+1) = n_1(t)\, \biggl [ \alpha\, f_1(t) + \biggl (1-\alpha\, f_1(t)\biggl )\biggl (\frac{q}{2}+ (1-q)\, f_1(t) \biggl ) \biggl ] + (N-n_1(t)) \bigl (1-\alpha(1-f_1(t))\bigl) \biggl(\frac{q}{2}+(1-q)f_1(t) \biggl ) \ .
\end{equation}
We now convert the previous relation into an equation for the increment $n_1(t + 1)-n_1(t)$. After a few simplifications, dividing by $N$ and taking the continuous limit ultimately delivers an equation for the time derivative $f'_1(t)$:
\begin{equation}\label{eq:twostatedynamics}
  f'_1(t) = (2\, f_1(t)-1)\,\bigl [ \alpha \, f_1(t)\, (1-f_1(t))\, (1-q)\, +\frac{q}{2}\, (\alpha-1)\bigl] \ . 
\end{equation}
After writing the corresponding equation for $f'_2(t)$, it can be shown that $f'_1(t)+f'_2(t) = 0$, as it should be. Therefore, the system is effectively described by just one differential equation. We choose the one in Eq.~\ref{eq:twostatedynamics}, we remove the index $1$ from $f$ and set $f'(t) = 0$ to find steady state solutions. This is a cubic equation, yielding three solutions $f_\infty = \lim_{t \rightarrow\infty} f(t)$ which can be expressed as functions of $q$ and $\alpha$. One solution is trivial and equal to the constant $1/2$, reflecting the fact that an initial condition in which $f_1(0) = f_2(0) = 1/2$ remains unchanged. The two non-trivial solutions read
\begin{equation}
    f^{\pm}_\infty = \frac{1}{2}\, \pm \frac{1}{2} \sqrt{\frac{q(\alpha-2)+\alpha}{\alpha(1-q)}},
\end{equation}
with $f^+_\infty+f^-_\infty = 1$, capturing the fact that each solution describes the popularity of one of the
two fields in the stationary state. Notably, from the above expression we can see that $f^\pm_\infty$ are real 
numbers only for $q \leq \alpha/(2 -\alpha)$, which we can then identify with the threshold $q^*$ in this simplified two-state setting. This, in fact, is the point where the constant solution $f^\pm_\infty = 1/2$ becomes the stable one. In Fig.~\ref{fig:analytical} we show the three steady state solutions as functions of $q$ for $\alpha = 1/2$, resulting in $q^*=1/3$. Notably, this holds despite the fact that such critical value is obtained under the assumption of a constant value for $\alpha$, which in this case is assumed to be equal to its average value, i.e., $\alpha = \langle \alpha \rangle = 1/2$. All in all, this highly stylised version of our model is sufficient to illustrate that a threshold separates a regime where actions are equally played from one where one action dominates, eventually resulting in  condensation when $q \rightarrow 0$.

\section*{Discussion}
Individuals and organisations are encouraged to imitate the behaviour of top performers in rankings. However, there is mounting evidence that this strategy is not the best one~\cite{performanceability,Giacomo2019,talentvsluck}. In this paper, we further support this assertion by considering an agent-based model to investigate the role of imitation and serendipity in social dynamics. In particular, we study the cumulative payoff that agents gain by choosing actions within a set based on these two mechanisms. We find that society reaches a higher payoff when imitation prevails, but at the cost of higher inequality. Moreover, in this scenario we find no correlation between an agent's accumulated payoff and their fitness. This is evident in the fact that the most skilled agents are not necessarily the most successful ones, clearly indicating that society is not meritocratic. 

We also examine how agents distribute themselves among the available actions. Our conclusion is that, when imitation is the predominant mechanism diversity is significantly reduced, as all agents tend to concentrate on one single action, which may not even be associated with meaningful societal benefits. On the contrary, when decisions are left to chance, outcomes become more serendipitous. In fact, society becomes more egalitarian, with a higher correlation between payoffs and skills, and a significant portion of available actions are played at any given time. These results are reminiscent of some of the findings published in~\cite{talentvsluck,challet2020origins}. However, let us remark that the driving mechanisms underpinning the two models are different, as we consider interactions between agents through imitation.

The two extremal conditions are separated by a sharp transition. Notably, sharp transitions have been observed in other social systems where agents tend to conform to one another, e.g., in social climbing phenomena~\cite{socialclimbinggame}, in the formation of political parties with a strong leader~\cite{HOLYST2023102137}, in stock markets~\cite{kirman1993ants} or even in the sudden emergence of traffic jams~\cite{kerner1997experimental}. This transition can be related to mechanisms which are similar to the ones responsible for phase transitions in physical systems~\cite{LEVY200571}. 

Our model is rather stylised, but at the same time extremely general, as we deliberately avoid to specify the details of the social system it describes. As such, its agents and their interactions capture a wide variety of situations. For instance, when the Covid-19 pandemic broke out it immediately became --- through imitation --- the main theme of research in most STEM fields, and in just a few months an incredible volume papers on this subject were published~\cite{covid}. Notably, the rush led to several documented errors in methodology and conclusions~\cite{covid2}. In fact, in a ``publish or perish''~\cite{de2005publish} environment incentives ---both at the level of authors and publishers ---
often favour condensation-like phenomena, i.e., the crowding towards certain research topics.

Switching to an entirely different domain, agents in our model may be interpreted as entrepreneurs starting a new business, needing to decide whether they should pursue the latest trend or something novel. The literature on the recent startup bubble shows that imitation-based entrepreneurial strategies lead to mixed results~\cite{tsolakidis2020impact}, which is in line with the outcomes produced by our model.

A possible extension of our model could incorporate time varying societal payoffs, i.e., allowing the $\pi_j$ parameters to be functions of time. This could capture the fact that the importance of certain actions could change due to exogenous events impacting society. For instance, going back to the aforementioned example, the sudden emergence of the Covid-19 pandemic obviously drove a massive change in importance of the research topics related to it. In this respect, our model captures scenarios in which the agents' decision-making evolves over time scales that are much shorter than those that characterise changes in the societal payoffs of actions.

\bibliography{bibliography}

\section*{Acknowledgements}
We acknowledge INFN, Sezione di Pavia, for the availability of computer resources which we used to perform simulations for the results reported in this paper.

\newpage

\appendix

\section*{Appendix}

All the simulations reported in this Appendix have been implemented with a different dynamics respect to the primary manuscript. Agents persevere in their current action with a probability proportional to $\alpha_{ij} \, f_j(t) \, \pi_j$, namely the switching is governed by the social impact as well. Parameters are $N=250$ agents, $M=100$ actions and $T=500$ time steps. Results are an average over a sample of size $S=1000$. 

\begin{figure}[h!]
\centering
\includegraphics[width=0.45\linewidth]{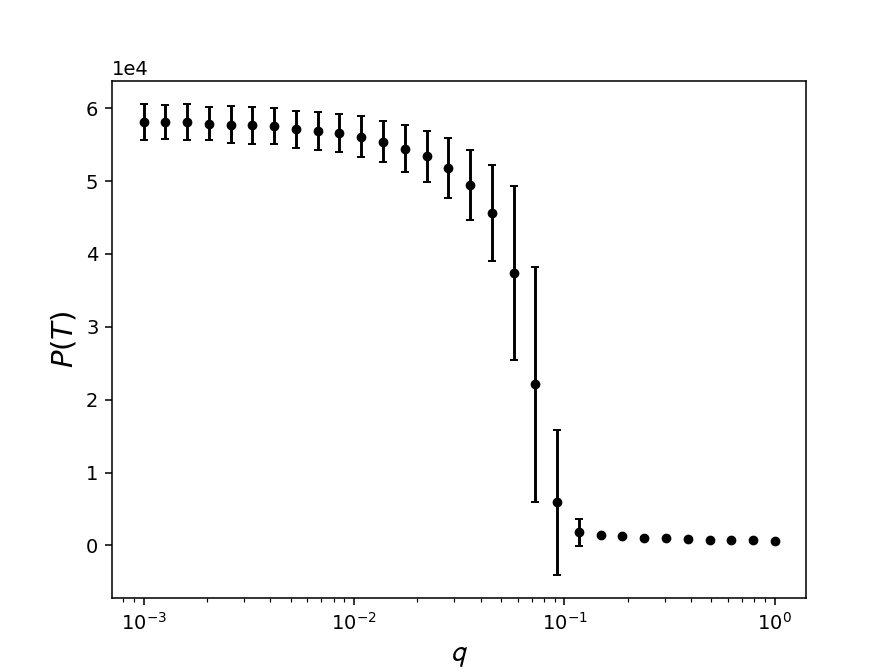} $\qquad \qquad$ \includegraphics[width=0.45\linewidth]{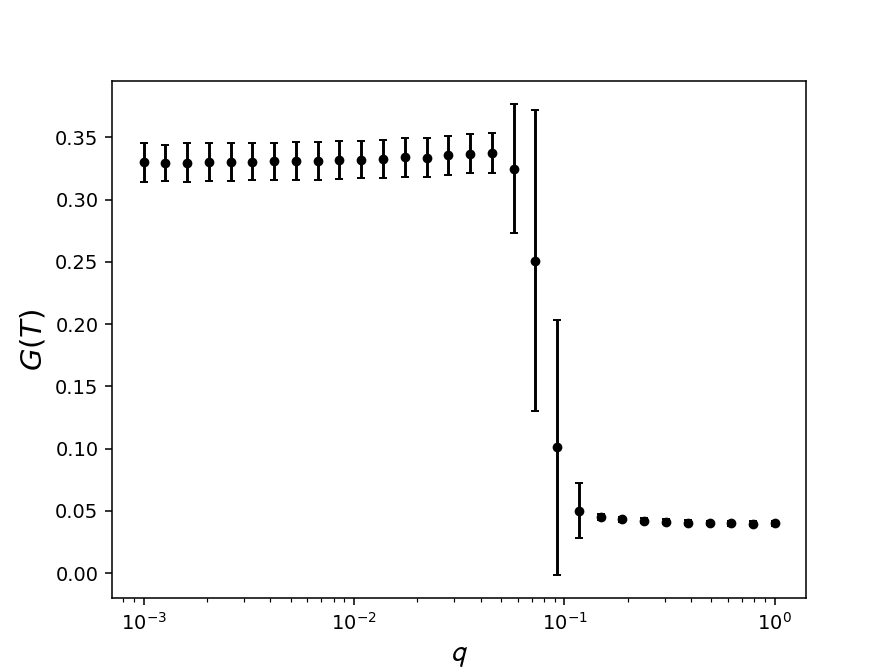}
\caption{Left panel: Cumulative societal payoff $P(T)$ at the end of the simulation for several values of parameter $q$. Right panel: Gini coefficient $G(T)$ for same values of $q$.}
\label{fig:payoff-gini_app}
\end{figure}

\begin{figure}[h!]
\centering
\includegraphics[width=0.45\linewidth]{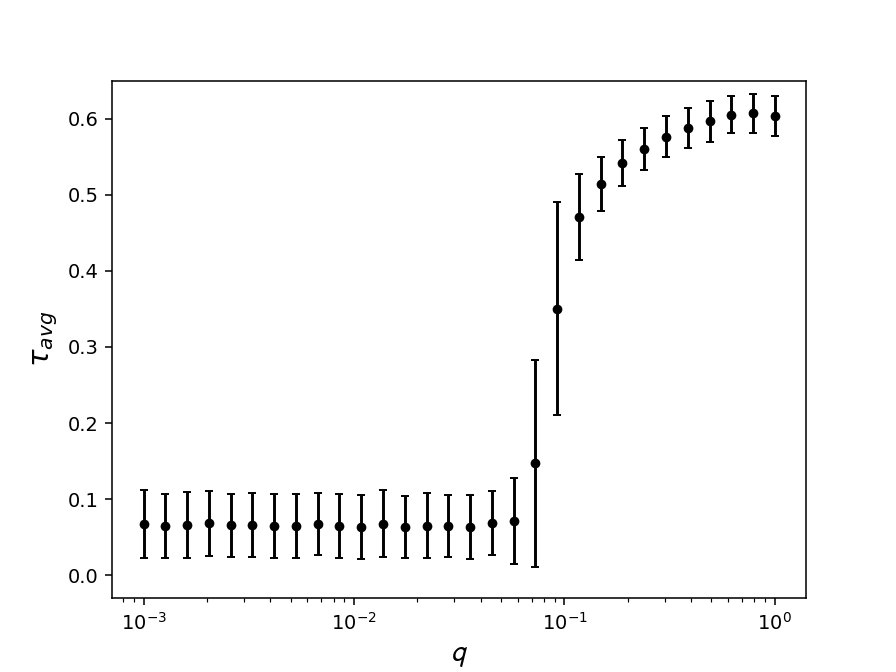} $\qquad \qquad$ \includegraphics[width=0.45\linewidth]{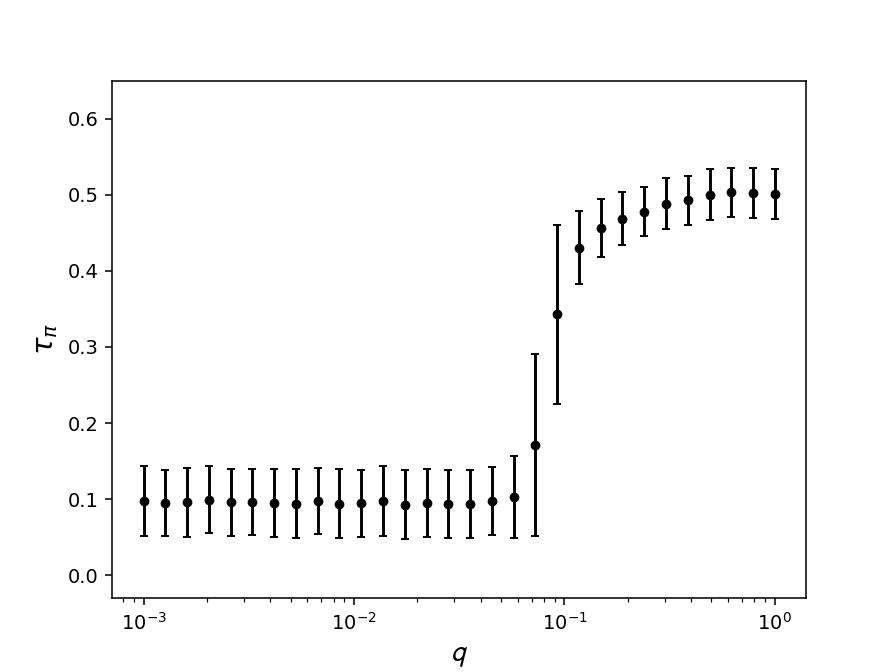} 
\caption{Kendall rank correlation coefficient $\tau$ between $P_i(T)$ and agent's fitness. Left panel: fitness is defined as in the main text, i.e. $\phi_i^{avg} = 1/M\, \sum_{j=1}^M \alpha_{ij}$. Right panel: a new definition of fitness is adopted, namely $\phi_i^{\pi} = 1/M\, \sum_{j=1}^M \alpha_{ij}\, \pi_j$}
\label{fig:correlation_app}
\end{figure}

\begin{figure}
\centering
\includegraphics[width=0.45\linewidth]{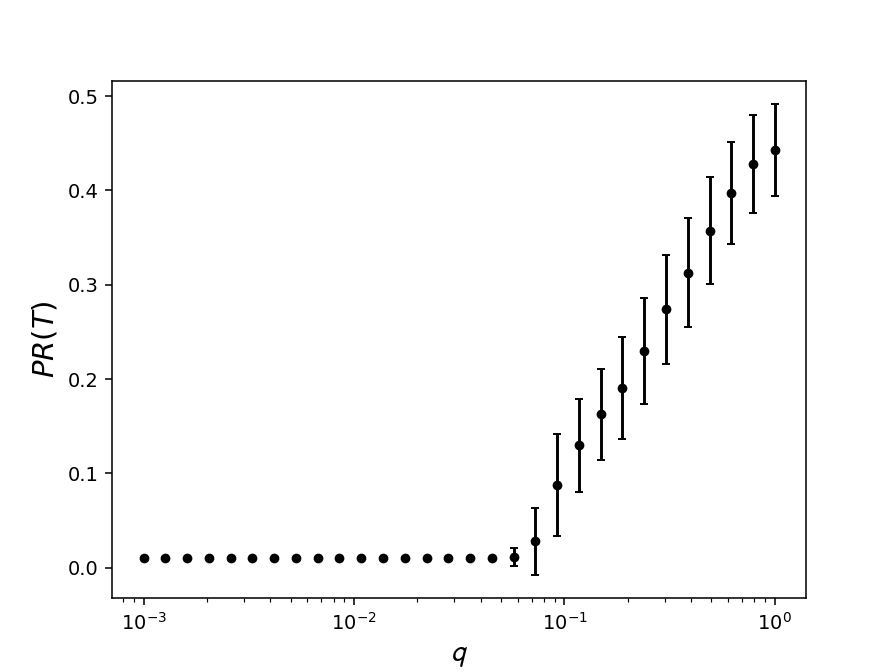} $\qquad \qquad$ \includegraphics[width=0.45\linewidth]{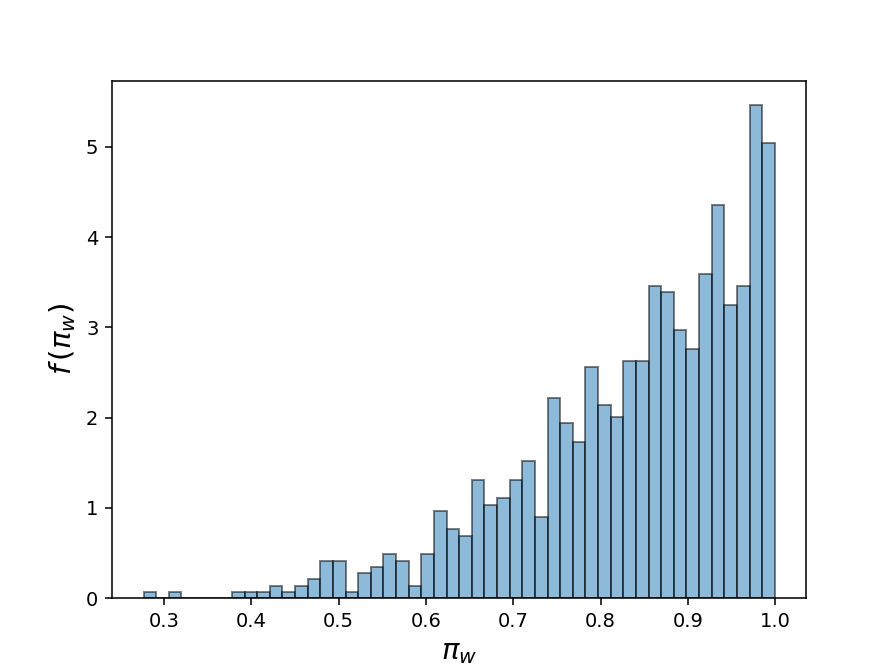}
\caption{Left panel: Participation ratio $PR(T)$. Right panel: normalized distribution of the payoffs of the most played actions at time $T$ for $q = 0$.}
\label{fig:ipr_app}
\end{figure}

\begin{figure}
\centering
\includegraphics[width=0.45\linewidth]{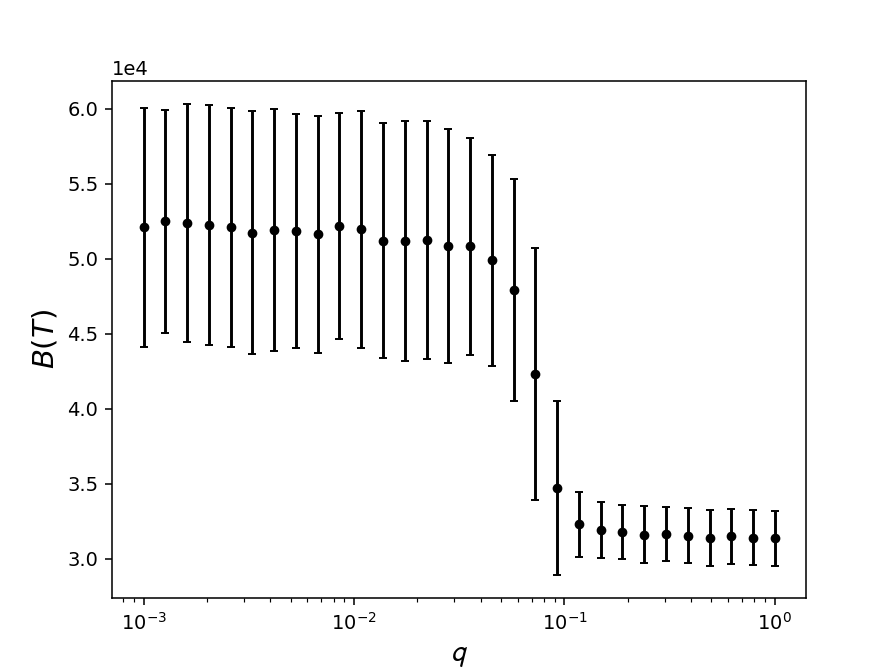}
\caption{The overall benefit $B(T)$ gained by the community due to agents' abilities in socially beneficial actions.}
\label{fig:benefit_app}
\end{figure}

\end{document}